\documentclass[12pt]{article}

\usepackage[dvipsnames]{xcolor}
\usepackage{amsmath}
\usepackage{amssymb}
\usepackage{dsfont}
\usepackage[english]{babel}
\usepackage[utf8]{inputenc}
\usepackage{times}
\usepackage{graphics}
\usepackage{graphicx}
\usepackage[T1]{fontenc}
\usepackage[official,right]{eurosym}
\usepackage{url}
\usepackage{eso-pic}

\title{Social Responsibility of Algorithms: \\ an overview}
\author{Alexis Tsoukiàs, \\ CNRS-LAMSADE, PSL, Université Paris Dauphine}
\date{}

\begin{document}

\thispagestyle{empty}

\enlargethispage*{8cm}
 \vspace*{-38mm}

\AddToShipoutPictureBG*{\includegraphics[width=\paperwidth,height=\paperheight]{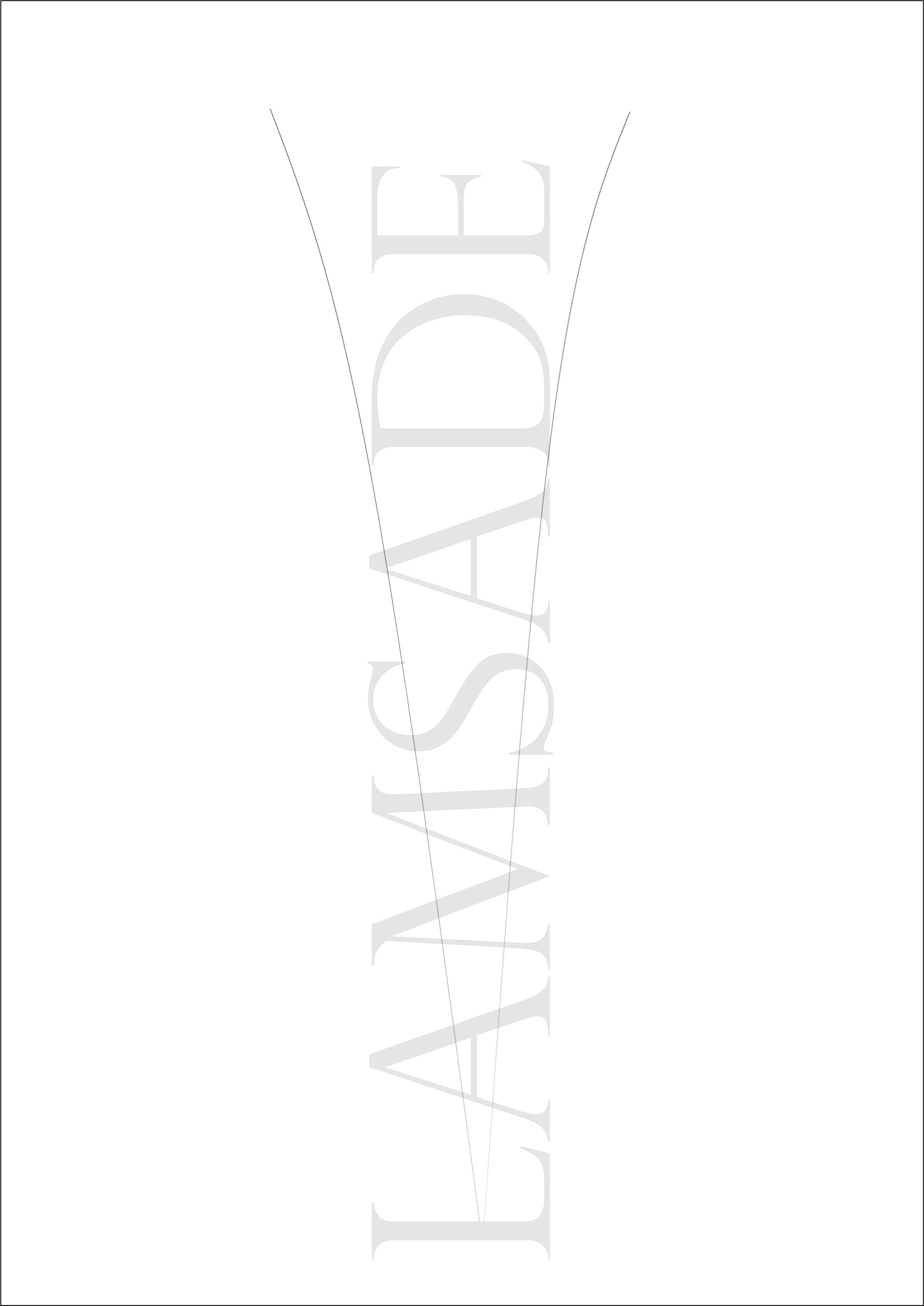}}

\begin{minipage}{24cm}
 \hspace*{-28mm}
\begin{picture}(500,700)\thicklines
 \put(60,670){\makebox(0,0){\scalebox{0.7}{\includegraphics{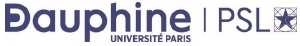}}}}
 \put(60,70){\makebox(0,0){\scalebox{0.3}{\includegraphics{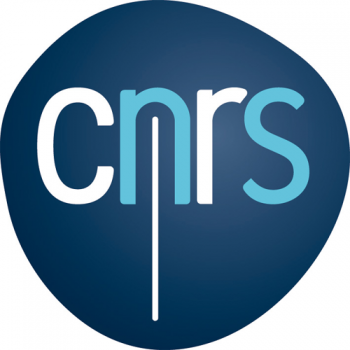}}}}
 \put(320,350){\makebox(0,0){\Huge{CAHIER DU \textcolor{BurntOrange}{LAMSADE}}}}
 \put(140,10){\textcolor{BurntOrange}{\line(0,1){680}}}
 \put(190,330){\line(1,0){263}}
 \put(320,310){\makebox(0,0){\Huge{\emph{397}}}}
 \put(320,290){\makebox(0,0){November 2020}}
 \put(320,210){\makebox(0,0){\Large{Social Responsibility of Algorithms:}}}
 \put(320,190){\makebox(0,0){\Large{an overview}}}
 \put(320,100){\makebox(0,0){\Large{Alexis Tsoukiàs}}}
 \put(320,670){\makebox(0,0){\Large{\emph{Laboratoire d'Analyses et Mod\'elisation}}}}
 \put(320,650){\makebox(0,0){\Large{\emph{de Syst\`emes pour l'Aide \`a la D\'ecision}}}}
 \put(320,630){\makebox(0,0){\Large{\emph{UMR 7243}}}}
\end{picture}
\end{minipage}

\newpage

\addtocounter{page}{-1}

\maketitle

\abstract{Should we be concerned by the massive use of devices and
algorithms which automatically handle an increasing number of
everyday activities within our societies? The paper makes a short
overview of the scientific investigation around this topic, showing
that the development, existence and use of such autonomous artifacts
is much older than the recent interest in machine learning
monopolised artificial intelligence. We then categorise the impact of
using such artifacts to the whole process of data collection,
structuring, manipulation as well as in recommendation and decision
making. The suggested framework allows to identify a number of
challenges for the whole community of decision analysts, both
researchers and practitioners.}

\newpage

\section{Motivations}

There is increasing concern around us about the impact of using automatic
devices making decisions for several aspects of our life, including credit
scoring, admissions to Universities, pricing of goods, recommender systems, up
to automatic vehicles or predictive justice (see \cite{AbuElyounes2020},
\cite{Citron2008}, \cite{Krolletal2017}). However, the use of algorithms in
order to automatise decision making is not recent (\cite{DavenportHarris2005});
actually algorithms exist even before computer science became the industry we
know. We can summarise the situation today under the following observations:

 \begin{itemize}
   \item We are creating and using autonomous artifacts with increasing
       decision autonomy.
   \item We have autonomous artifacts with increasing learning capacities.
   \item There is evidence of biased decisions, of counterintuitive
       decisions, of inappropriate use of personal and sensible data, of
       unforeseen consequences, when such devices are largely adopted and
       used\footnote{The best known controversy is the ``COMPASS'' case:
       \url{https://www.propublica.org/article/machine-bias-risk-assessments-in-criminal-sentencing}}.
   \item Software editing and data services are concentrated to few
       industrial players.
 \end{itemize}

The aim of this paper is to clarify a number of issues which affect both
researchers and practitioners interested in decision support (decision
analysts). It turns out that many of the concerns we are discussing today,
already existed in the literature (see for instance \cite{WinogradFlores86})
and are less ``new'' and ``urgent'' from what they appear to be. On the other
hand, the extension today of designing, testing and actually using autonomous
artifacts represents a real challenge for the community of decision analysts.
The paper aims at identifying which are these challenges and how can we
appropriately handle them.

The paper is structured as follows. In Section 2 we make a brief survey of the
literature with no pretention to be exhaustive, essentially in order to
identify the principal trends. Section 3 introduces the principal concepts
through which we can establish a common framework. Section 4 presents two brief
examples which help understanding the topics discusses in th previous section.
Finally, Section 5 summarises the challenges we have in front of us the next
years.

\section{Historical background}

The literature about Decision Support Systems dates back to the 70s: see the
seminal paper \cite{GoryScottMorton1971} and the two well known books
\cite{KeenScottMorton78} and \cite{SpragueCarlson82}. This literature builds
upon already existing research and practice with ``Management Information
Systems'' (see \cite{MasonMitroff73}). The idea is simple: exploit the
information existing and circulating within an organisation in order to improve
decision making under different types of requirements (see also the interesting
discussion in \cite{LandryPascotBriolat83}).

In more recent days the same idea came alone under the concept of ``analytics''
(or business analytics or business intelligence; see \cite{DavenportHarris07}).
The ``new'' idea is to extend the use of data in order to support decision
making creating and assessing massive data bases (more or less open), thanks to
a large increase of computing capacity. However, the application of these ideas
remains bounded at supporting ``human decision makers'' within organisations,
the scope of ``analytics'' being to produce suitable information for decision
makers.

A relatively innovative idea has been instead to increase the decision capacity
of ``autonomous artifacts'' in order to improve the overall performance of
complex systems. However, once again automatising decisions is not a totally
new idea; we can see how this evolved through the following topics.

\begin{itemize}
  \item Automatically conducted vehicles have been designed since a century
      ago: automatic pilots for aircrafts date at the beginning of the 20th
      century (see \cite{AstromMurray2008} or \cite{StevensLewis1992}).
      Automatically controlled devices and robots exist since the middle of
      the 20th century (\cite{Hunt1985}, \cite{Wiener48}) and represent today
      a very important scientific and industrial area.
  \item Multi-Agent systems started being designed in the 80s (see
      \cite{RusselNorvig95} or \cite{Wooldridge2002}) allowing software
      agents to perform with increasing decision autonomy.
  \item Recommender Systems appeared soon after as software platforms where
      consumers could be automatically guided among huge catalogs of goods
      and being advised about their choices matching their preferences with
      products features and the behaviour of similar consumers (see
      \cite{Aggarwal2016} or \cite{Riccietal2011}).
  \item Blockchains introduced the possibility to decentralise trust
      construction procedures through distributed cryptography on the web
      (see \cite{Nakamoto2008}, \cite{Narayananetal2016}).
\end{itemize}

As can be noted the idea of increasing the decision capacity of autonomous
artifacts already has several decades of development, including commercial and
industrial applications of large scale (virtually any aircraft today is
automatically driven and most e-commerce platforms include a recommender
system). There have been though two breakthroughs: \\
 - the increasing availability and accessibility to data (of any type and
 quality, including personal and sensible ones); \\
 - the massive expansion of ``deep learning algorithms'' allowing high level
 correlations among data with excellent accuracy and predictive capacity (for a
 presentation see \cite{Goodfellowetal2016}, while for an interesting
 discussion about correlation and causality see \cite{Pearl2009}).

Such developments fuelled a literature about the impact and the consequences of
automated decision making. This literature has been essentially focussed around
three areas.

\begin{enumerate}
  \item \textbf{Fairness}. Since the seminal paper \cite{Dwork2012}, there
      have been several tentatives in order to establish a general definition
      of ``fairness'' for decisions taken by algorithms. This notion of
      fairness assumes the existence of ``protected'' groups within the
      society, which are potentially threaten by biased algorithmic decision
      making processes (see also \cite{HajanDomingoFerrer2013} and
      \cite{Leprietal2018}). However, such ``protected groups'' are only
      recognised within certain countries and it soon appeared that there are
      several formal and substantial difficulties in establishing a model of
      general validity (see \cite{Friedleretal2016}).
  \item \textbf{Accountability and Explicability}. Not independent from the
      fairness issue there has been the discussion about the accountability
      of algorithms (see \cite{STUD2019} and \cite{Wieringa2020}). The issue
      here is the possibility to provide convincing explanations on why an
      algorithm would end taking a certain type of decisions (possibly
      unfair, biased or counterintuitive). The topic includes explicability
      of data mining and machine learning algorithms (see
      \cite{Guidottietal2018}) with specific emphasis to the case where the
      algorithms behave as black boxes with unpredictable behaviour (such as
      deep neural networks).
  \item \textbf{Ethics}. Finally there has been discussion about the ethical
      dimension of automated decision making. The issue arises essentially in
      the case of automatically conducted and/or unmanned vehicles and
      devices which may need to take decisions with high ethical impacts
      (such as impacting human life: see \cite{Bonnefonetal2016} and
      \cite{Rahwan2018}). The topic however, has gone beyond this specific
      area questioning the possibility and/or opportunity to endow algorithms
      with ethical principles (see for instance \cite{Greeneetal2016}).
\end{enumerate}

The result of such discussions has been the creation of new scientific
communities, possibly interdisciplinary ones, the largest for the moment being
the ACM-FAccT series of conferences (see \\
\url{https://facctconference.org/index.html}).

\section{What is the problem?}

The survey presented in the previous section, far from being exhaustive,
highlights the fast growth of an area of scientific investigation, but also of
public concern. In reality there exist several different problems which both
scientific paper and press and blogs tend to put together under different
``titles'' basically sharing a number of keywords: Artificial Intelligence,
Data Protection, Algorithmic Transparency etc. (see \cite{Leprietal2017},
\cite{ONeil2016}). Most of them tend to raise concerns of the general public of
how such technologies could impact our life. It pays, however to clarify a
number of issues starting with establishing precisely the object of scientific
investigation.

From our perspective this object is the \textit{``design, implementation and
systematic use of autonomous artifacts with enhanced decision capacity}. In the
following we are going to analyse which are the different problems this object
includes.

In conducting our analysis we will adopt an industrial production perspective
because we are talking about the evolution of an industry whose raw material
are data. Under such a perspective we are going to focus upon the raw material
itself (the data), the transformation process (the algorithms), the
implementation (the software), the outcome and the impact to the society.
However, before analysing the components of this industrial process we may
analyse a number of fundamental topics.

\subsection{Fundamentals}

The first fundamental topic to remember is that automation is not a
straightforward perspective, but a choice. There are plenty of examples around
us of processes which are not automated and nobody thinks to automatise them.
If automation is a choice then there is somebody who makes the choice and there
should be reasons for which this choice has been done. Automation for certain
types of production has been decided by the industry and their management
essentially in order to increase profits (although several times quality of the
products has been used as a reason). Automation of certain industrial processes
has been decided for safety purposes or in order to alleviate workers from
unhealthy or dangerous activities. If automatising a decision process is a
choice, we should always ask ourselves who decides to automatise, for which
reasons and who is going to pay the cost of it. If the process to automatise
concerns the public (such as college admissions or predictive justice) there is
a matter of democracy and citizens' participation to such decisions.

The second fundamental topic to remember is that decisions imply responsibility
and responsibility implies liability for the consequences of any decision. Each
time we consider automatising a decision process we should always ask ourselves
who is liable for the decisions taken by the autonomous artifact we create. In
the flying industry this issue has been long time solved: liable are the
airlines who use aircrafts with automatic pilot and there is a chain of
responsibilities, certifications and training in order to keep such liability
clear. The liability issue does not concern solely the principle, but also the
practical aspect: be sure that responsibilities can be traced, recognised and
affected to those who could be liable. Automatising a decision process means
that we have a clear idea of how the liability issue is going to be considered.

The third fundamental topic concerns the fact that algorithms can mirror how
our societies are, but cannot change them. It is clearly annoying discover
through what an algorithms learns that our societies are unfair, discriminate
minorities, behave aggressively, in other terms are politically incorrect. But
these are the societies as our democracies shaped them. If we do not like them,
there are democratic paths for changing our societies, but algorithms will
always mirror what our societies actually are. We cannot introduce innovation
in society just designing innovative algorithms.

\subsection{The raw material}

The raw material of the type of processes we are concerned are data. Data are
collected, stored, retrieved and manipulated and each single step of these
processes could have an impact upon the whole decision process to automatise.
There are two basic topics to consider as far as the use (term resuming all the
above steps) of data is concerned.

The first topic concerns the rights an individual (a citizen) and/or a group
have upon certain data. Data (of any type) do not belong specifically to
somebody and for certain types of data we could consider them as ``commons''.
However, we can have certain rights upon certain data and as soon as these
rights are established we can consider whether these can be traded. However,
trading rights implies establishing clear contracts. The problem today is that
there is an absolute information asymmetry (see \cite{Mas-Colelletal1995})
between each single citizen and his rights on the one side and the data
industry on the other side. Besides, there is a value scaling about data
availability: the value of the rights I have upon my personal data alone is an
extremely small fraction of the value of owing the rights of millions of
individuals.

The second topic concerns the certification of the data used within automated
decision processes. Biased data will result in biased outcomes. Noisy data will
result in bad quality outcome. Corrupted data will result in unverifiable
outcomes. There is necessity to certify the whole pipeline of collecting,
storing and retrieving data used for any automated decision process (see
\cite{Christophidesetal2020}).

\subsection{The outcome}

First of all we need to make an important distinction. Autonomous artifacts can
provide two types of outcomes: ``decisions'' and ``recommendations''. For this
purpose we may define a decision as an \textit{irreversible allocation of
resources to tasks or actions}. In the first case is the artifact that makes
such an allocation which results in some action being undertaken, while in the
second case the artifact only makes a recommendation (generally to a human
agent) which ``decides''.

From a practical point of view there are very few autonomous artifacts which
actually have full decision autonomy and generally this concerns ``low level''
actions in automatic controlled devices (such as in self-conducted vehicles).
Most of the automated decision processes concern in reality artifacts which
suggest a certain action to be undertaken. It can be the case of credit
scoring, of predictive justice scores, college admissions, job candidates
screening etc.. However, this ``final decision freedom'' of the human agent is
far from being a warranty about the controllability of the final outcome. Most
automatically formulated recommendations are rarely contested and usually are
followed by the human decision makers, which essentially explains why such
\textit{suggestions} are regularly considered as \textit{decisions}. In the
following we will focus on automated ``recommendation'' processes, since these
are the most frequent (and complex).

A first issue to consider is the fact that the result of information
manipulation is never straightforward: there is no (and will never exist)
universal procedure through which we can obtain from raw data a synthesis. Data
manipulation ought being \textit{meaningful} (see \cite{roberts79}),
\textit{useful} (see \cite{Thebook00}) and \textit{legitimate} (see
\cite{Tsoukias07aor}), these requirements still allow for plenty different
procedures. It is a matter of choice for the designers and users.

The second issue, following from the previous one, is that we may desire adding
further properties to the outcome: we may desire having a recommendation which
is \textit{fair, unbiased, neutral etc.}. The fact is that there is no unique
definition to such concepts. Both economists in mechanism design theory
(\cite{HurwiczReiter2006}, \cite{Maskin2008}) and computer scientists more
recently (\cite{Friedleretal2016}) realised that there are several different
ways to define notions such as ``fairness'', each corresponding to different
hypotheses about the society, the inequalities within the society and the ways
to prevent or to correct them. This means we need to establish both the
requirement of a feature to meet and a formal definition for each requirement
and how to test it.

Establishing which requirements the recommendations needs to meet is a matter
of choice. The third issue is to know who decides which requirements an outcome
of a given autonomous artifact have to be satisfied. Several of such
requirements might be inconsistent one with respect to another. Somebody (who?)
has to make a choice resulting in satisfying a certain property and thus,
failing to satisfying another one. Under such a perspective it is important
when designing an autonomous artifact to know which
properties/requirements/axioms an automated recommendation procedure satisfies
and which not. This is rarely the case today (the reader can check that no
recommender systems specifies how notes are aggregated among users and products
and thus nobody knows which properties are satisfied by such procedures).

What happens in case the autonomous artifact is ``data driven'': in other terms
the outcome depends essentially upon the data feeding process, but the data
manipulation is unknown (as happens for many black-box automated procedures)?
The fourth issue related to the quality of the outcome concerns the ``hidden
values'' embedded within many autonomous artifacts. Decisions and
recommendations are never based directly on raw data. Between these and any
decision there are ```preferences'' or ``values'' which allow to compute a
``choice'' (or whatever else a decision or a recommendation may mean; see
\cite{ColorniTsoukias2013}). Preferences and values are always subjective and
represent an individual or a society of individuals. If an autonomous artifact
is able to make a decision or to compute a recommendation it means that
somebody embedded within the artifact his/her preferences. And these are
independent from how the artifact turns to learn out from the data feeding it.
It turns out that is of paramount importance to know how values are actually
embedded in any of such systems and/or how these are learned (see
\cite{FurnkranzHullermeier2010}).

\subsection{The process}

It is often the case that not only the outcome of a process matters, but also
the process itself. This is both the case for automated decisions and automated
recommendations. The former might need to be explained, justified, tested and
proven to be ``correct'' in case of accidents, misbehaviour, unforeseeable
consequences etc. The latter might need to be trusted, defended, argued,
recused, might need to be convincing, trustworthy, understandable, etc.. In all
such cases we need to check whether the autonomous artifact is
\textit{accountable}. However, there are several different levels of
\textit{accountability}.

  \begin{enumerate}
   \item Given an algorithm or to be more precise a bundle of algorithms
       setting an automatic decision procedure can we trace precisely what
       these algorithms do?
   \item Provided that we can trace the execution of the algorithms, can we
       provide ``explanations'' (interpretable, understandable, usable) to
       any type of stakeholder about the choices done and the obtained
       results?
   \item Provided we can trace and explain the behaviour of an algorithm can
       we provide the ``ultimate reasons'' for which the algorithm/automatic
       device made a precise decision or recommendation? If it is the case
       can we replicate the decision providing the same input?
   \item Supposing the algorithm cannot guarantee replicability (for instance
       in case the algorithm learns each time is executed we cannot guarantee
       that for a given input the output will remain the same) what type of
       explanations/justi\-fi\-cations/reasons would be considered satisfying
       in case of a dispute?
  \end{enumerate}

Besides the above introduced aspects of accountability there are also long term
consequences to take into account when a certain type of autonomous artifact is
largely used in the real world. How should we define accountability for the
long term impact of e-commerce platforms using recommender systems (using
certain types of algorithms) for promoting their business?

\subsection{The implementation}

Autonomous artifacts are essentially software. Certainly in the case of robots
and other autonomous devices there are physical parts which are equally
important, but the essential of what we are talking is software. Indeed
algorithms and procedures not necessarily are implemented in software, but we
are concerned with the ones who actually are used under form of computer
programs.

The first issue to consider is the formal verification that a given software
implementation of a bundle of algorithms endowing an autonomous artifact,
actually does what these algorithms are expected to do. This is far from being
self-evident and the more complex the artifact is, the more difficult the
verification becomes.

The second issue concerns security. Any software implementation can be attacked
and/or manipulated. We can certainly choose safer, redundant and highly
protected implementations (as the stakes of the artifact scope increase; see
the case of e-voting), but this comes at a price which needs to be
commensurable to the benefits and the value of the automation.

The third issue concerns the use of open source software. While this apparently
could be inconsistent with straight security requirements, open source software
remains the ultimate possibility to analyse why an autonomous artifact actually
acts as observed. While security issues can easily be handled even when using
open source code, being able to check the code through collective intelligence
processes remains a fundamental warranty for most accountability issues.

\subsection{The impact to society}

Introducing a drug to a living system has expected and unexpected consequences.
It is exactly for this reason that new drugs before being cleared and allowed
to be used are extensively tested under rigid protocols, are permanently
checked and submitted to scrutiny and possibly can be retired from commerce.
Usually there is an independent authority which takes care of this process. We
are going to use this ``metaphor'' (introducing a new drug to a living system)
in order to consider the long term impact of introducing an autonomous artifact
in handling some aspect of our everyday life.

Should we demand a certification for the whole process (raw material, outcome,
process and implementation) before allowing to release an autonomous artifact
in the society? Should we create an independent agency or authority for this
purpose? While many of the known unknowns can be handled through appropriate
design and preliminary testing, the only way to discover the unknown unknowns
is to do extensive testing and monitoring.

The use of autonomous artifacts for some types of business implies modifying
the business model of the enterprise and/or the organisation introducing this
``innovation''. The issue is whether, stakeholders, consumers, users are aware
of the consequences such a modification will introduce upon the goods and
services delivered by that type of business. Organisational studies are plenty
of innovation failure cases (\cite{RhaiemAmara2019}) for businesses,
organisations and markets, because the process was poorly designed, not
understood, not fitting the expectations, undesired etc. and this includes the
choices about automatising decisions and processes.

The industry of automated decisions and recommendations is dominated by few big
players, both with respect to the collection, storage, retrieval and use of
data and services and with respect to software editing and engineering.
Monopolies never benefited consumers and this case will not be an exception.
This market, as many others, needs regulations and these need to be global.

If we need to audit autonomous artifacts and monitor their long term impact and
if we need to establish global rules for this market we need to bear in mind
that the life cycle of these products can be short (even very short) compared
to the length of audits and regulations. It might make sense to be innovative
as far as the timing of regulation is concerned if we do not want to miss any
real opportunity to control.

\section{Examples}

The following examples are voluntarily not among the typical ones used within
the Artificial Intelligence literature just in order to show that several
issues discussed in this paper are far beyond the AI challenges.

\subsection{Automatic pricing}

Automatic pricing became popular since the late 70s because of the innovation
introduced by American Airlines: yield management. The simple idea consists in
adjusting prices and seats offered on commercial flights following the demand
prediction and possible capacity of the airline (see \cite{Smithetal1992}).
Today is a regular practice, not only for the airlines industry: many retailers
practice automatic pricing in order to optimise revenue management. That said,
we can make a number of observations.

 \begin{enumerate}
   \item Implementing the automatisation of this activity has been a choice,
       both for profit maximisation purposes and for gaining competitive
       advantages for the first runners. It is less obvious whether this
       resulted in better and less cheap services for the customers. In any
       case it was not a choice of the users who can find exactly the same
       product at several, significantly different, prices (see
       \cite{Mauri2007}).
   \item There exist several different economic models helping to compute
       automatically prices, depending upon the type of product, the type of
       the market and the hypotheses about the consumers' behaviour (see
       \cite{CrossR1997}). It is actually unknown whether the choice of any
       among such models has been discussed before using them.
   \item We know instead that adopting a precise model of pricing,
       considering the density of the competition (on the retailers market),
       can lead to unforeseen consequences, as in the famous ``Stapler''
       case\footnote{\url{https://www.wsj.com/articles/SB10001424127887323777204578\\189391813881534}},
       where the same object (a stapler) was sold at different prices in
       different neighbours. The use of competition density resulted in
       discounting the object in the ``rich neighbours'' (high density) and
       sell it at full price in the ``poor neighbours'' (low density). Not
       necessarily this was the policy and will of the retailer.
   \item Automatic pricing strongly depends upon the quality and timing of
       the necessary data feeding the economic and decision models of yield
       management. However, there is no warranty that the data pipeline for
       any of the retailers adopting such tools is reliable and trustworthy.
       This is all the more important in case part of the data feed a
       black-box learning procedure for which we may have no convincing
       justifications available. On the other hand, the liability for any
       ``wrong'' decision remains internal to the retailer who will just have
       to absorb the consequences in their business.
   \item Automatic pricing modified how the travel industry is organised and
       influenced how people travel and organise their leisure time. In other
       terms it had a huge impact upon the whole society (as often happens
       when industries introduce new products or services). On the other hand
       what is the long term impact of such new patterns of mobility and
       leisure time consumption? Are these sustainable at a long run? Nobody
       ever discussed that, when yield management models and algorithms have
       been introduced (more than 40 years ago).
 \end{enumerate}

\subsection{Voting}

Voting is not an automatic decision procedure, or at least is not perceived as
such. However, the reader will note that when we adopt the term voting we
implicitly consider voting procedures (algorithms) which ``compute'' a
``winner'' of an electoral contest. There exist several such algorithms and
generally they may yield totally different results even when the preferences of
the voting society are clear. The fact is that we (the society) need to make
choices of which among such algorithmes should be used and possibly we (the
society) have to trust that the result is legitimate. For a presentation of
electoral systems see \cite{Reynoldsetal2005}, while a theoretical
investigation about social choice theory can be found in \cite{arrow1book51} or
in \cite{Sen86}. For an interesting survey about such methods being considered
under a computational aspect see \cite{Brandtetal2016}. Once again we make a
number of observations.

 \begin{enumerate}
   \item We vote in order to elect representatives, presidents, committees,
       chairs etc. and this is done for legitimating governing. However,
       ``electing'' is not the only way to appoint representatives,
       committees, chairs ... Our societies (after centuries of struggles)
       decided to use such procedures (which might result in less efficient
       decision procedures, but certainly more legitimated). We vote because
       we want to.
   \item As already mentioned there exist many different voting procedures
       and algorithms computing the winner(s). It is well known that it does
       not exist and it will never exist an universal procedure, because even
       simple ``democratic'' requirements are inconsistent among them and
       cannot be satisfied simultaneously. This means we need to choose one.
       In doing that it pays knowing which requirements are satisfied and
       which not and this has been the scope of large part of the social
       choice theory literature.
   \item Different voting procedures promote different views about our
       societies, the ways to govern the society and about citizens'
       participation (see \cite{Rae1971}). Moreover, each of such systems
       need to make choices on how ``fair representation'' should be
       interpreted (proportionality among citizens, among regions, among
       ethnic groups are typical topics which are typically impossible to
       satisfy all together). These are political choices which need to be
       discussed as such and not as technical problems.
   \item Electronic vote is increasingly popular, but has been tested to be
       easily hacked, corrupted and manipulated, while manual procedures are
       far more complicated to alter (at least under usual democratic
       institutions operating). As already noted previously, the software
       version of an algorithm does not coincide with the algorithm itself.
 \end{enumerate}

\section{Conclusion}

Let us try to summarise our overview. Does it make any sense to talk about the
\textit{social responsibility of algorithms}? Technically speaking, no, since
algorithms cannot be liable for what they compute. Designers, clients demanding
for algorithms, software editors can be considered responsible (and thus,
liable), but not the algorithms. On the other hand, the use of algorithms in
order to improve our decision making is older that computer science itself and
the demand for extending their use, for creating further autonomous artifacts
with decision capacity is never lasting.

As decision analysts we share part of the responsibility of how such autonomous
artifacts are shaped, designed, implemented and used in the real world. Under
such a perspective we should pay attention and further develop our theoretical
research as well as our reflection about our practices around the following
topics: \\
 - characterising algorithms and procedures through the properties they satisfy
 or do not satisfy; \\
 - remembering that each time we choose a precise procedure in order to solve a
 given decision problem this is rarely a straightforward choice, but one among
 many options and as such needs to be justified and considered for its impact
 beyond that precise problem; \\
 - characterising and specifying the data to be used by algorithms, reflecting
 the three basic requirements: meaningfulness, usefulness and legitimacy; \\
 - analyse how our methods, procedures and protocols are used and adopted
 within real organisations and within our societies.

Algorithms and formal models will never stop being used in order to improve how
decisions are taken both by humans and machines. It is upon the designers
define what improvement means and for whom. This is our social responsibility.

\section*{Acknowledgements}

This document follows discussions which took place during the two workshops
about this topic in Paris, December 2017 and 2019
(\url{www.lamsade.dauphine.fr/sra}) and I am indebted to the participants for
their contributions. Although several ideas are due to these discussions, I
remain the sole responsible for this essay.

\bibliographystyle{plain}
\bibliography{complete-bibliography,temp}

\end{document}